\documentstyle[12pt,psfig]{article}
\makeatletter
\def\eqalign#1{\null\vcenter{\def\\{\cr}\openup\jot\m@th
  \ialign{\strut$\displaystyle{##}$\hfil&$\displaystyle{{}##}$\hfil
      \crcr#1\crcr}}\,}
\makeatother
\begin{document}
%\hfill {\small November 1998}
\bigskip
\bigskip
\bigskip
\begin{center}
{\Large\bf 
Eigenvalue density for a class of Jacobi matrices\footnote{
Dedicated to the memory of the late Professor V. I. Peresada} 
}\\
\bigskip
\bigskip
\bigskip
\bigskip
{\large I. V. Krasovsky}\\
\bigskip
Max-Planck-Institut f\"ur Physik komplexer Systeme\\
N\"othnitzer Str. 38, D-01187, Dresden, Germany\\
E-mail: ivk@mpipks-dresden.mpg.de\\
\medskip
and\\
\medskip
B.I.Verkin Institute for Low Temperature Physics and Engineering\\
47 Lenina Ave., Kharkov 310164, Ukraine.
\end{center}
\bigskip
\bigskip
\bigskip

\noindent
{\bf Abstract.}
We obtain the asymptotic distribution of eigenvalues of real symmetric
tridiagonal matrices as their dimension increases to infinity and  
whose diagonal and off-diagonal elements asymptotically change with
the index $n$ as $J_{nt+i\;nt+i}\sim a_i\varphi(n)$,
$J_{nt+i\;nt+i+1}\sim b_i\varphi(n)$,  $i=0,1,\dots,t-1$, where
$a_i$ and $b_i$ are finite, and $\varphi(n)$ belongs to a certain class of
nondecreasing functions.

\newpage
A very versatile method for calculation of physical properties of quantum
systems ([\ref{P1}--\ref{KP}] and references in [\ref{SSP}--\ref{KP}]) 
uses the Lanczos procedure to reduce 
the matrix of the Hamiltonian or some other relevant operator to the 
tridiagonal form. After this reduction, various properties of tridiagonal
(Jacobi) matrices provide much help in solving the problem at hand. 
An important problem
arising in connection with this method is to find the asymptotic 
distribution of
eigenvalues of Jacobi matrices (when the dimension of the matrix increases
to infinity) if the asymptotics of the matrix elements is known. 
The purpose of this note is to present a solution to this problem in a 
particular case. 

Let $J$ be a symmetric Jacobi matrix with real matrix elements (that is 
$J_{ik}=J_{ki}$, $i,k=0,1,2,\dots$, $J_{ik}=0$ if $|i-k|>1$,
$J_{i\;i+1}\ne 0$ ). Suppose that
$\lim_{n\to\infty}J_{nt+i\;nt+i}/\varphi(n)=a_i$ and
$\lim_{n\to\infty}J_{nt+i\;nt+i+1}/\varphi(n)=b_i$, $i=0,1,\dots,t-1$,
where $a_i,b_i<\infty$ are not all simultaneously zero;
$t$ is a positive integer;
the function $\varphi(x):{\bf R}^+\to{\bf R}^+$
is nondecreasing and satisfies the condition
$\lim_{n\to\infty}\varphi(n+x)/\varphi(n)=1$ for any real $x$.
Note that for $\varphi(n)=1$ the matrix $J$ is asymptotically periodic.

Furthermore, let $J(n)=||J_{ik}||_{i,k=0}^{nt-1}$ denote the 
$nt\times nt$ truncated $J$. Our aim is to calculate the eigenvalue
density $\rho(z)$ of $J(n)/\varphi(n)$ as $n\to\infty$. 
It is defined by the requirement that $\rho(z)dz$ be 
in the limit $n\to\infty$ the relative number of eigenvalues of
$J(n)/\varphi(n)$ in the interval $dz$.

We now impose a further
limitation on the form of $\varphi(n)$: suppose that there exists a continuous
function defined by $g(\omega)=du/d\omega$, where 
$\omega(u)=\lim_{n\to\infty}\varphi(nu)/\varphi(n)$ for $u\in[0,1]$. 
Note that for  $\varphi(n)=n^\gamma$, $\gamma> 0$, we have
$g(\omega)=\omega^{-1+1/\gamma}/\gamma$.

The ideas of these definitions come from the works [\ref{Nev},\ref{Ull}],
where the case $t=1$ was considered, and
the moments of the eigenvalue density $\rho(z)$ 
were calculated. The explicit expression for $\rho(z)$
was given in [\ref{Ull}] for $t=1$, $a=0$. It is called Nevai-Ullman 
distribution. Using a different approach, Van Assche has obtained for 
$\varphi(n)=n^\gamma$, and arbitrary $a$, $b$ the density $\rho(z)$ for $t=1$
[\ref{VA1}] and (implicitly) for $t=2$ [\ref{VA2}].  
We will now obtain $\rho(z)$ for any $t$ and $a$, $b$ by generalizing the
method of [\ref{Nev},\ref{Ull}].

Before that, let us recall some facts about periodic symmetric Jacobi matrices.
Let $L$ be such a matrix defined by the formulas
$L_{nt+i\;nt+i}=a_i$ and $L_{nt+i\;nt+i+1}=b_i$, $i=0,1,\dots,t-1$,
$n=0,1,\dots$. The same $a_i$, $b_i$ appear in our definition of $J$.
Define two systems of polynomials $p_k(x)$ and $q_k(x)$ by the initial
conditions $p_0=1$, $p_1=(x-a_0)/b_0$; $q_0=0$, $q_1=-b_{t-1}/b_0$ and
the same (for $y_k=p_k$ or $y_k=q_k$) recurrence relation:
\[
y_{k+1}=((x-a_k)y_k-b_{k-1}y_{k-1})/b_k,\qquad k=1,2,\dots,t-1.
\]
Let $S(x)=p_t(x)+q_{t-1}(x)$. The spectrum of $L$ consists of exactly $t$
bands: the image of $[-2,2]$ under the inverse of the transform
$S(x)=\lambda$. The boundaries of the bands (denote them $\mu_i$, $\nu_i$,
$i=1,\dots,t$,
$\nu_0\equiv-\infty<\mu_1\le\nu_1\le\mu_2\le\nu_2\le\cdots\le
\nu_t<\mu_{t+1}\equiv\infty$) 
are solutions to the equations $S(x)=\pm 2$. Thus, the spectrum
of $L$ is ${\rm spec}(L)=\cup_{i=1}^t[\mu_i,\nu_i]$. 
The eigenvalue density of $L$ is given by the formula
$\rho_0(x)=|dS(x)/dx|/(t\pi\sqrt{4-S(x)^2})$, 
for $x\in{\rm spec}(L)$, and $\rho_0(x)=0$ otherwise.

Note that for $t=1$ we have $\rho_0(x)=(\pi\sqrt{4b_0^2-(x-a_0)^2})^{-1}$.

{\bf Theorem}
{\it
Let $f(z,\omega)=g(\omega)\rho_0(z/\omega)/\omega$.
Then, under the conditions specified above,
the eigenvalue density of $J(n)/\varphi(n)$ as 
$n\to\infty$ is the following:
 
1) $0\in[\mu_k,\nu_k]$ for some $k$ ($1\le k\le t$),
\[
\rho(z)=\cases{
\left[\int_{z/\nu_i}^1+\sum_{j=i+1}^t\int_{z/\nu_j}^{z/\mu_j}\right]
f(z,\omega)d\omega, 
& $z\in[\mu_i,\nu_i]$, $\mu_i\ge 0$\cr
{\it the\;preceding\;expression\;with\;\it i=k}, & $z\in[0,\nu_k]$\cr
\left[\sum_{j=i+1}^t\int_{z/\nu_j}^{z/\mu_j}\right]f(z,\omega)d\omega, 
& $z\in(\nu_i,\mu_{i+1})$, $\nu_i\ge 0$
\cr
\left[\int_{z/\mu_i}^1+\sum_{j=1}^{i-1}\int_{z/\mu_j}^{z/\nu_j}\right]
f(z,\omega)d\omega, 
& $z\in[\mu_i,\nu_i]$, $\nu_i\le 0$\cr
{\it the\;preceding\;expression\;with\;\it i=k}, & $z\in[\mu_k,0]$\cr
\left[\sum_{j=1}^i\int_{z/\mu_j}^{z/\nu_j}\right]f(z,\omega)d\omega, 
& $z\in(\nu_i,\mu_{i+1})$, $\mu_{i+1}\le 0$
};
\]

2) $0\in(\nu_k,\mu_{k+1})$ for some $k$ ($0\le k\le t$),
The expression for $\rho(z)$ is the same as above
(two formulas for $z\in[0,\nu_k]$ and $z\in[\mu_k,0]$
should be dropped) except in the interval
$(\nu_k,\mu_{k+1})$ where
\[
\rho(z)=\cases{
\left[\sum_{j=k+1}^t\int_{z/\nu_j}^{z/\mu_j}\right]f(z,\omega)d\omega, 
& $z\in(0,\mu_{k+1})$\cr
\left[\sum_{j=1}^k\int_{z/\mu_j}^{z/\nu_j}\right]
f(z,\omega)d\omega, 
& $z\in(\nu_k,0)$.}
\]
}

\bigskip

{\it Sketch of the proof.}
Some basic ideas of the proof we borrow from [\ref{Nev},\ref{Ull}]
where the case $t=1$ was considered. 
Let us denote $K_M=\int_{-\infty}^\infty x^M \rho_0(x) dx$, where $M$ is
a nonnegative integer, and verify first the following identity:
\begin{equation}
\lim_{k\to\infty}{1\over t}\sum_{i=0}^{t-1}
\left(\left[{J\over\varphi(k)}\right]^M
e_{kt+i},e_{kt+i}\right)=K_M,\label{i1}
\end{equation}
where $\{ e_j\}_{j=0}^\infty$ is an orthonormal basis in which the above
defined matrices of operators $J$ and $L$ are written.
Indeed, it is obvious that the l.h.s. of this identity is equal to
\[
\lim_{k\to\infty}{1\over t}\sum_{i=0}^{t-1}(L^Me_{kt+i},e_{kt+i}),
\]
while
\begin{equation}
\lim_{k\to\infty}{1\over t}\sum_{i=0}^{t-1}(L^Me_{kt+i},e_{kt+i})=
\lim_{n\to\infty}{1\over nt}\sum_{j=0}^{nt-1}(L^Me_j,e_j)=K_M.
\end{equation}
From (\ref{i1}) it follows that
\begin{equation}
{1\over t}\sum_{k=0}^{n-1}\sum_{i=0}^{t-1}(J^Me_{kt+i},e_{kt+i})=
\sum_{k=0}^{n-1}\varphi(k)^M(K_M+\epsilon_k),\label{i2}
\end{equation}
where $\epsilon_k\to 0$ as $k\to\infty$.
We have:
\begin{equation}
\lim_{n\to\infty}\frac{\sum_{k=0}^{n-1}\varphi(k)^M}
{n\int_0^1\varphi(xn)^Mdx}=1,
\qquad
\lim_{n\to\infty}\frac{\sum_{k=0}^{n-1}\epsilon_k\varphi(k)^M}
{\sum_{k=0}^{n-1}\varphi(k)^M}=0.
\end{equation}
Hence, (\ref{i2}) can be rewritten as follows:
\begin{equation}
{1\over t}\sum_{k=0}^{nt-1}(J^Me_k,e_k)=
(K_M+\delta_n)n\int_0^1\varphi(xn)^Mdx,\label{i3}
\end{equation}
where $\delta_n\to 0$ as $n\to\infty$.

To proceed further, we will need to estimate the sum
\begin{equation}
Z_n(M)=\sum_{k=0}^{nt-1}(J^Me_k,e_k)-
\sum_{k=1}^{nt}x_{k,n}^M,
\end{equation}
where $x_{k,n}$, $k=1,\dots,nt$ are the eigenvalues of $J(n)$.
Note that $Z_n(M)$ is the difference between traces of two matrices, one of
which is obtained by first taking the power $J^M$ 
and then truncating the result; and the other one, by first truncating
$J$ and then taking the power $J(n)^M$. 
It is easy to calculate that for $n$ large enough $Z_n(0)=Z_n(1)=0$, 
$Z_n(2)=J_{nt\;nt-1}^2$, $Z_n(3)=J_{nt\;nt-1}^2(J_{nt\;nt}+J_{nt-1\;nt-1})$, 
etc. In general, we have for all sufficiently large $n$:
\begin{equation}
|Z_n(M)|\le f(M)\max_{0\le i,k\le n+M}|J_{i\;k}|^M,
\end{equation}
where $f(M)$ does not depend on $n$.
In view of this estimate, (\ref{i3}) leads to the identity:
\begin{equation}
\lim_{n\to\infty}\frac{\sum_{k=1}^{nt}x_{k,n}^M}{nt\int_0^1\varphi(xn)^Mdx}=
K_M.\label{i4}
\end{equation}

Changing the variables $x_{k,n}=\varphi(n)z_{k,n}$ and taking the limit,
we have:
\begin{equation}
\eqalign{
m_M\equiv\lim_{n\to\infty}{1\over nt}\sum_{k=1}^{nt}z_{k,n}^M=
\int_0^1\omega^M(x)dx K_M=
\int_0^1\omega^M g(\omega) d\omega
\int_{-\infty}^\infty x^M \rho_0(x) dx,\\
M=0,1,\dots,}\label{m}
\end{equation}
where $\omega(x)$ and $g(x)$ are defined as above.

Changing the variables $z=\omega x$, $\psi=\omega$ in the
double integral in (\ref{m}), we get:
\begin{equation}
m_M=\int_{-\infty}^\infty z^M\rho(z) dz,\label{m2}
\end{equation}
where $\rho(z)$ is defined in cases 1 and 2 as in Theorem.
Since the actual support of integration in (\ref{m2}) is finite, the 
function $\rho(z)$ is defined uniquely by $m_M$, $M=0,1,\dots$. 
Thus, $\rho(z)$ can be identified with the eigenvalue density. 

\bigskip

Although the simple case $t=1$ has appeared to a large extent 
in the literature, we will now discuss it to give the reader a better
feeling of the results and to show the relation to other methods
to investigate the distribution of eigenvalues. 

So from now on let $t=1$ and define $a_0=a$, $b_0=b/2$.  
Henceforth, we put $a\ge 0$ and $b\ge 0$ without loss of generality
(it is easily seen that $\rho(z)$ for the pairs $a$, $b$
and $a$, $-b$ is the same). 
It will be interesting to note the distinction between the cases 
$a>b$ and $a<b$.

\noindent{\bf Corollary of the theorem} {\it
Let $t=1$, $a_0=a$, $b_0=b/2$.
Under the conditions specified above,
the eigenvalue density of $J(n)/\varphi(n)$ as $n\to\infty$
is the following: 

1) $0\le a/b \le 1$, 

\noindent
$\rho(z)=
\frac{1}{\pi}\int_{z/(a+b\,{\rm sign}\,z)}^1\frac{g(\omega)d\omega}
{\sqrt{b^2\omega^2 -(z-a\omega)^2}}$  if  $z\in[a-b,a+b]$,
and $\rho(z)=0$ otherwise;

2) $a/b>1$,

\[
\rho(z)=\cases{
\frac{1}{\pi}\int_{(a-b)/(a+b)}^1\frac{g(\omega\frac{z}{a-b})d\omega}
{\sqrt{b^2\omega^2-(a-b-a\omega)^2}}, 
& $z\in[0, a-b]$\cr
\frac{1}{\pi}\int_{z/(a+b)}^1\frac{g(\omega)d\omega}
{\sqrt{b^2\omega^2 -(z-a\omega)^2}}, & $z\in[a-b, a+b]$\cr
0, & $z\notin[0,a+b]$}.
\]
}

Note that for $\varphi(n)=n^\gamma$, $\gamma>0$ we can write case 2
in a simpler form:
\[
\rho(z)=\cases{h(a-b)\left(\frac{z}{a-b}\right)^{-1+1/\gamma}, 
& $z\in[0, a-b]$\cr
h(z),& $z\in[a-b, a+b]$\cr
0, & $z\notin[0,a+b]$},
\]
where
\[
h(z)=
\frac{1}{\pi\gamma}\int_{z/(a+b)}^1\frac{\omega^{-1+1/\gamma}d\omega}
{\sqrt{b^2\omega^2 -(z-a\omega)^2}}.
\]

The form of $\rho(z)$ is illustrated for $\varphi(n)=n^\gamma$, 
in Figures 1,2. When $\gamma\to 0$ the density $\rho(z)$ for all $a,b$
approaches the distribution $\rho_0(z)=(\pi\sqrt{b^2-(z-a)^2})^{-1}$,
$z\in[a-b,a+b]$.

\begin{figure}
\centerline{\psfig{file=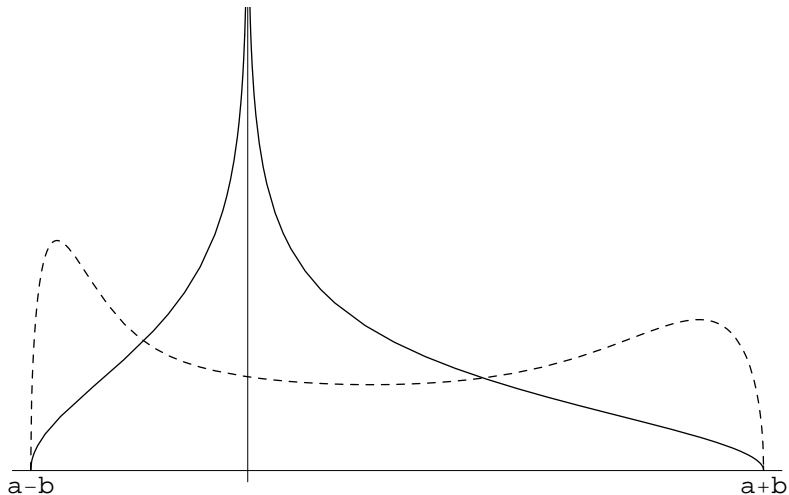,width=5in,angle=0}}
\vspace{0.5cm}
\caption{Example of the density of eigenvalues $\rho(z)$ 
for $a<b$, $\varphi(n)=n^\gamma$. Solid curve 
corresponds to some $\gamma\ge 1$, dashed curve, to $0<\gamma<1$. } 
\label{f1}
\end{figure}
\begin{figure}
\centerline{\psfig{file=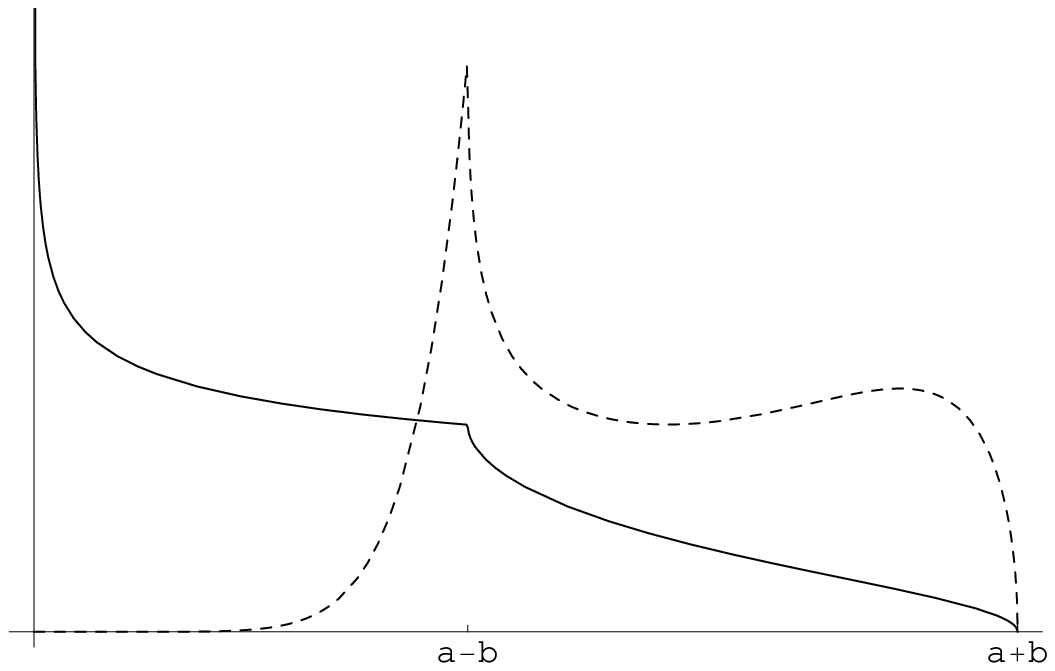,width=5in,angle=0}}
\vspace{0.5cm}
\caption{Example of the density of eigenvalues $\rho(z)$ for  
$a>b$, $\varphi(n)=n^\gamma$. Solid curve corresponds to
some $\gamma>1$, dashed curve, to $0<\gamma<1$ (in fact, it is obvious
from the picture that the second derivative of the dashed curve is positive
for $z\in[0,a-b]$, so it corresponds to some $\gamma<1/2$).} 
\label{f2}
\end{figure}

One can note that for $\varphi(n)=n$ the matrix $J/r$, where
$r=\sqrt{|a^2-b^2|}$, has the same asymptotics of the matrix elements
as the matrix associated  with the Meixner (for $a>b$) or 
Meixner-Pollaczek (for $a<b$) orthogonal polynomials.\footnote{
Here we disregard a sign in front of the matrix and polynomials.} 
Asymptotics of the eigenvalues of $J(n)/r$ is the
asymptotics of the zeros (to the main order) of these polynomials.

For $\varphi(n)=n$ the function  $\rho(z)$ is elementary.
Integration gives in cases 1 and 2, respectively
\[
1)\quad
\rho(z)=\cases{
\frac 1{\pi r} {\rm arccosh}\left|{r^2\over zb}+{a\over b}\right|, 
& $z\in[a-b,a+b]$\cr
0, & $z\notin[a-b,a+b]$};
\]
\[
2)\quad
\rho(z)=\cases{1/r, & $z\in[0, a-b]$\cr
\frac 1{\pi r} {\rm arccos}\left(-{r^2\over zb}+{a\over b}\right), 
& $z\in[a-b, a+b]$\cr 0, & $z\notin[0,a+b]$},
\]
\[
0\le\arccos(x)\le\pi.
\]
Replacing $r$ by 1 in these formulas gives
the contracted asymptotic densities of zeros of the 
Meixner-Pollaczek (case 1) and Meixner (case 2) polynomials.
The quantities $a$, $b$ are expressed in this case in terms of the parameters
of the polynomials. For detailed discussion of zero densities for various
systems of orthogonal polynomials see [\ref{KvA}] and references therein.

The Meixner and  Meixner-Pollaczek polynomials satisfy difference equations
of the type [\ref{Koe}]:
\begin{equation}
B(x)p_n(x+\delta)-C(x,n)p_n(x)+D(x)p_n(x-\delta)=0.
\end{equation}
By analyzing such equations, it is possible to obtain information on the
asymptotic distribution of zeros of $p_n(x)$ [\ref{adzdde}]. Thus,
corollary of the theorem for $\varphi(n)=n$ can be given an alternative 
proof, and also more precise information (than the density) about asymptotics 
of zeros can be found: distribution 
of zeros of the Meixner polynomials $p_n(nz)$ in the region $z\in[0,a-b]$
where $\rho(z)=1$.  
\bigskip
\bigskip

\noindent
{\large\bf Acknowledgements}
\medskip

\noindent
I thank Alphonse Magnus and Walter Van Assche for useful 
correspondence.

\end{document}